\documentclass[9pt,twocolumn,twoside]{opticajnl}
\journal{opticajournal} 

\setboolean{shortarticle}{true}

\DeclareCaptionLabelSeparator{custom}{\textbar}
\DeclareCaptionFormat{custom}{\textsf{\textbf{#1 #2} \small #3}}
\newcommand{\reffig}[2]{Fig.~\ref{#1}\textbf{\textsf{#2}}}

\definecolor{Myblue}{cmyk}{0.8,0.6,0,0.1}
\newcommand{\change}[1]{{#1}}

\usepackage{lineno}

\title{Laser diode self-injection locking to an integrated high-Q Fabry-P\'erot microresonator}

\author[1]{Alexander E. Ulanov}
\author[1]{Thibault Wildi}
\author[1,2]{Utkarsh Bhatnagar}
\author[1,2,*]{Tobias Herr}

\affil[1]{Deutsches Elektronen-Synchrotron DESY, Notkestr. 85, 22607 Hamburg, Germany}
\affil[2]{Physics Department, University of Hamburg UHH, Luruper Chaussee 149, 22761 Hamburg, Germany}

\affil[*]{tobias.herr@desy.de}

\begin{abstract}

Self-injection locking (SIL) of laser diodes to microresonators is a powerful technique that enables compact narrow linewidth lasers. Here, we extend this technique to chip-integrated Fabry-Perot microresonators, which offer high-quality factors and large mode volumes in a compact footprint, reducing fundamental thermo-refractive noise. The resonators consist of a silicon nitride waveguide terminated by two photonic crystal reflectors fabricated via scalable ultraviolet lithography. Evanescent side-coupling allows precise tuning of the SIL feedback mechanism. We present a theoretical model and experimentally demonstrate SIL, resulting in a fundamental thermorefractive-noise-limited laser. We achieve excellent agreement between the experimental results and theoretical model. These results complement current SIL techniques and are relevant to chip-scale low-noise laser systems.
\end{abstract}

\setboolean{displaycopyright}{false} 

\begin{document}

\maketitle

\section{Introduction}

Many advanced laser applications, such as atomic clocks \cite{chou:2011}, long-range sensing \cite{martins:2022}, high-resolution spectroscopy \cite{xu:2024}, quantum communications \cite{gisin:2007}, and metrology \cite{degen:2017}, benefit from low-noise and narrow-linewidth lasers. Such lasers can be implemented using various methods, including locking to an external cavity, but they typically require complex setups. For many emerging applications, more compact and lower complexity systems are desirable.

An attractive approach that can meet these criteria is self-injection-locking (SIL) of semiconductor laser diodes (LD) to external high-quality factor $Q$ resonators \cite{kondratiev:2023a}. SIL relies on a narrow linewidth resonant optical feedback from a high-Q resonator that is re-injected into the LD, causing frequency locking and a collapse of the LD's linewidth.
Since early demonstration with tilted Fabry-P\'erot (FP) cavities \cite{dahmani:1987}, SIL has been explored with fiber \cite{wei:2016} and bulk FP-cavities \cite{lewoczko-adamczyk:2015, liang:2023}, crystalline \cite{vassiliev:1998} whispering-gallery-mode (WGM) and photonic chip-integrated ring-type microresonators with internal backscattering \cite{corato-zanarella:2023} or external reflection \cite{shim:2021}. 
Recent demonstrations based on chip-integrated ring-type resonators include Hertz-linewidth \cite{jin:2021} and frequency-agile semiconductor lasers \cite{siddharth:2024}, ultralow noise electrical signal generators \cite{liang:2015, savchenkov:2024}, battery-operated \cite{stern:2018}, turnkey integrated soliton microcomb sources \cite{shen:2020}, that can be implemented to operate deterministically based on synthetic reflection SIL \cite{ulanov:2024} and with metrology-grade performance \cite{wildi:2024}. A comprehensive review on this topic can be found elsewhere \cite{kondratiev:2023a}.

A limitation of chip-integrated SIL systems is thermo-refractive noise (TRN), which is a direct consequence of their small optical mode volume \cite{huang:2019}. In this context, recently demonstrated chip-integrated high-$Q$ FP resonators \cite{yu:2019b, ahn:2022, wildi:2023} are attractive as they can provide access to large mode volumes in a simple linear geometry, complementing the established spiral and racetrack resonators.
However, achieving a suitable SIL feedback is non-trivial in a chip-integrated FP-resonator, as resonant back-reflection is absent. To achieve SIL in an FP resonator, the transmitted resonant wave must be looped back into the laser diode \cite{wei:2016, hao:2021, liang:2023, ousaid:2024}; this adds complexity, and the looped-back SIL signal may easily be disturbed or even overwhelmed by reflection at the FP's input coupling mirror.

Here, we demonstrate for the first time SIL of an LD to a chip-integrated FP microresonator. This resonator is built from two identical photonic crystal reflectors (PCRs) connected by a straight waveguide section \cite{wildi:2023}. To create a SIL feedback, we couple to the FP cavity not through the PCR but instead utilize a directional evanescent coupler approaching the FP cavity from the side \cite{kazarinov:1987}. In this way, resonant light propagating in the FP cavity is extracted and fed back to the diode, with a magnitude that can be precisely controlled via the coupling strength. To describe the dynamics in this novel architecture, we present a theoretical model and, based on this model, experimentally demonstrate a low-frequency noise SIL laser limited by fundamental TRN and with prospects for further improvements.

\section{Theoretical model and simulations}

\begin{figure}[ht]
  \centering
  \includegraphics[width=\columnwidth]{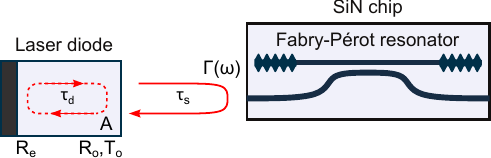}
  \caption{
      \textbf{Fabry-Pérot SIL.} Laser diode self-injection locking to a photonic chip integrated side-coupled Fabry-P\'erot microresonator.
            }
    \label{fig:FP_SIL_concept}
\end{figure}

For the theoretical model, we consider a semiconductor LD coupled to a chip-integrated FP microresonator using evanescent side-coupling as illustrated in \reffig{fig:FP_SIL_concept}{}. We consider only one spatial mode for the LD cavity, microresonator, and bus waveguide and assume perfect mode-overlap. Generally, the LD emits at a frequency $\omega$, different from $\omega_d$ and $\omega_m$, the closest LD cavity and FP microresonator eigenfrequencies, respectively. \change{For the scope of this paper, we will limit the analysis to the steady-state case; a description of the dynamical model can be found elsewhere \cite{tronciu:2017}. We assume a simplified model of the LD cavity formed by a highly reflective end mirror with reflection coefficients $R_e$ and an out-coupling mirror with reflection and transmission coefficients of $R_o$ and $T_o$, respectively (see \reffig{fig:FP_SIL_concept}{}).}

\change{Injection into the LD cavity of an external field leads to a modification of its net gain $n'_g = n_g + \delta$ and a refractive index change $n' = n + \alpha \delta$, where $\alpha$ is the Henry factor that quantifies the amplitude-phase relation of the gain medium. In the new steady state, the constant field amplitude $A$, which describes the field of frequency $\omega$ inside the LD is
\begin{equation} 
       A =  (i T_o)^2 \frac{A}{R_o} \Gamma(\omega) \, e^{i\omega \tau_s} + R_o R_e A \, e^{i\frac{\omega}{c} (n' - i n'_g) 2L} 
\label{eq:field_A}
\end{equation}
where $\Gamma(\omega)$ denotes the complex reflection coefficient of the microresonator, $\tau_s$ is the roundtrip time from the diode to the microresonator and back, and $L$ is the length of the LD. Taking into account that, for the free-running laser, gain compensates losses such that $R_o R_e \, e^{\omega_d n_g 2 L / c} = 1$, one can approximately solve \eqref{eq:field_A} for the frequency difference $\omega - \omega_d$ by expanding the small argument exponential and separating it into real and imaginary parts to obtain~\cite{kondratiev:2017}:
\begin{align}
     & \omega - \omega_d = \frac{1}{\tau_d}\frac{T_o^2}{R_o} \sqrt{1+\alpha^2} \left( \text{Im}\Gamma(\omega) \cos\psi + \text{Re}\Gamma(\omega) \sin\psi\right)
     \label{eq:SIL1}
\end{align}
where $\psi = \omega \tau_s - \mathrm{arctan} \alpha$ is the injection phase and $\tau_d$ is the LD cavity round trip time.} 
So far, no assumptions have yet been made about the physics of the reflector. In the case of a high-$Q$ side-coupled FP microresonator close to its eigenfrequency $\omega_m$ the complex reflection spectrum is given by
\begin{align}
    \label{eq:fp_complex_refl}
    \Gamma(\omega) &= \frac{\kappa_\mathrm{ex}}{\kappa/2 + i (\omega_m - \omega)}
\end{align}
where --- different from ring resonators --- the total loss rate follows $\kappa = \kappa_0 + 2 \kappa_\mathrm{ex}$,  with $\kappa_0$ and $\kappa_\mathrm{ex}$ being the microresonator intrinsic and coupling loss rates respectively.

Substituting \eqref{eq:fp_complex_refl} into \eqref{eq:SIL1}, we obtain the equation often referred to as the SIL tuning curve:
\begin{align}
     & \xi = \zeta + K_\mathrm{FP} \frac{\mathrm{sin}\psi - \zeta \mathrm{cos}\psi}{1 + \zeta^2}
     \label{eq:fp_SIL}
\end{align}
where $\xi = \frac{2}{\kappa}(\omega_m - \omega_d)$ is the normalized detuning of the free-running (without SIL) LD emission frequency from the nearest microresonator resonance $\omega_m$ and $\zeta = \frac{2}{\kappa}(\omega_m - \omega)$ is the normalized {\it effective} detuning of the emission frequency $\omega$ from the same microresonator resonance in SIL operation. 
\change{Further, $K_\mathrm{FP} = \frac{4 \eta}{\kappa} \frac{1}{\tau_d}\frac{T_o^2}{R_o} \sqrt{1+\alpha^2}$} is a combined coupling coefficient between LD and microresonator, where $\eta = \kappa_\mathrm{ex}/\kappa$ is the coupling coefficient of the FP microresonator. 
The injection phase $\psi$ is a free parameter that can be readily controlled experimentally, as described below. 

\begin{figure}[ht]
  \centering
  \includegraphics[width=\columnwidth]{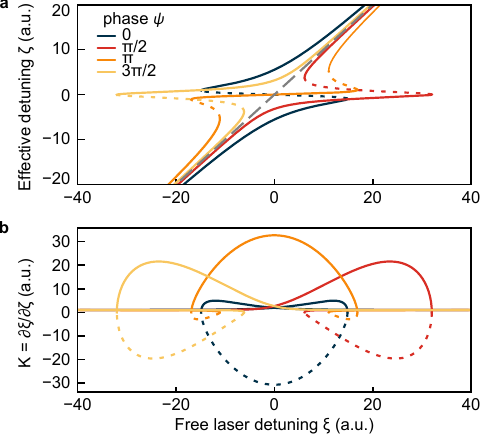}
  \caption{
      \textbf{SIL tuning curves.} Laser diode self-injection locking frequency tuning curves (a) and corresponding $\partial \xi / \partial \zeta$ slopes (b) for different injection phases.
      The lines are calculated using \eqref{eq:fp_SIL} and \eqref{eq:der_fp_SIL} for a Fabry-P\'erot microresonator with parameters $\eta = 0.2$, $\kappa/2 \pi = 100$ MHz, coefficient $K_\mathrm{FP} = 32$, and a short laser-to-reflector roundtrip time such that $\kappa \tau_s \ll 1$.
            }
    \label{fig:tuning_curves}
\end{figure}

The slope of the SIL tuning curve defines the stabilization coefficient $K = \partial \xi / \partial \zeta$, \change{which quantifies the ratio between free-running frequency fluctuations to those in the SIL state} (assuming fluctuations in $\omega_m$ are negligible), such that the laser linewidth is effectively reduced by a factor $K^2$ \cite{kondratiev:2017}.    
In \reffig{fig:tuning_curves}{}, both the SIL tuning curve and the stabilization coefficient $K$ are illustrated for different injection phases $\psi$. Unstable branches are represented by dashed lines. These curves show that for a fixed coefficient $K_\mathrm{FP}$, there exists an optimal injection phase ($\psi = \pi$) such that the SIL stabilization effect is maximized.

From a practical perspective, operating somewhat detuned from the resonance center is often beneficial. This is, for example, required to access the dissipative Kerr soliton existence range \cite{herr:2012}. With this in mind, we analyze the $\partial \xi / \partial \zeta$ derivative of \eqref{eq:fp_SIL}:
\begin{align}
     & K = 1 + K_\mathrm{FP} \frac{(\zeta^2 - 1)\mathrm{cos}\psi - 2\zeta \mathrm{sin}\psi}{(1 + \zeta^2)^2}
     \label{eq:der_fp_SIL}
\end{align}
For any detuning $\zeta$, there exists a range of $\psi$ such that the fraction on the right-hand side of \eqref{eq:der_fp_SIL} is positive. This indicates optimal stabilization is achieved when $K_\mathrm{FP}$ is maximized. For side-coupled FP-resonators, this is achieved when \change{$\eta = 0.25$ (in contrast to ring resonators where this is achieved for the critically coupled case $\eta = 0.5$)}. Since the denominator on the right-hand side of \eqref{eq:der_fp_SIL} scales with the $4^\mathrm{th}$ power of $\zeta$, it remains beneficial to operate at a low effective detuning $\zeta$. Finally, we note for clarity that we have not included Kerr-nonlinear effects in our treatment, \change{as they remain negligibly small at the experimentally used power levels. However, those can be readily included in analogy to previous work in ring resonators~\cite{voloshin:2021}, taking into account the counter-propagating light fields in an FP resonator~\cite{obrzud:2017a, cole:2018a}.}

\section{Experiments}

The FP microresonator is commercially fabricated via wafer-level ultraviolet lithography on a silica-clad SiN platform with an 800~nm layer thickness. It consists of two identical photonic crystal reflectors (PCRs) with a straight waveguide \change{(1.6~$\mu$m width)} section between them \cite{wildi:2023}. The PCRs are implemented through sinusoidal corrugation of both sidewalls of the straight waveguide with a period \change{of 480~nm,} design peak-to-peak corrugation amplitude of 500~nm, \change{and overall length of 200~$\mu$m.}
Coupling to the FP resonator from the bus waveguide is achieved via a directional evanescent coupler approaching the microresonator from the side \change{with a minimum gap of 500~nm} (Fig.~\ref{fig:FP_SIL_concept}). The coupling strength $\kappa_\mathrm{ex}$, and with it, the SIL feedback strength, can be tuned by adjusting the gap between the coupler and the resonator waveguides.  Specifically, we use an FP resonator with a free spectral range (FSR) of 20.34~GHz, corresponding to an optical length of approximately 3.87~mm. The total linewidth is $\kappa / 2 \pi=80$~MHz ($Q$-factor $Q=2.4$~million), and the coupling efficiency is $\eta = 0.19$, close to the theoretical optimum of 0.25. 

To explore SIL with the FP microresonator, we butt-couple an off-the-shelf semiconductor distributed feedback (DFB) LD emitting at 1558~nm to the photonic chip, achieving an on-chip power of 2.5~mW. The DFB LD is mounted on a 3-axis piezo translation stage, which permits adjusting the injection phase by slightly modifying the gap between LD and microresonator chip; alternatively, tuning of the injection phase can be accomplished in a fully integrated setting via integrated microheaters or electro-optic phase modulators. Transmitted light is outcoupled from the chip with an ultra-high numerical aperture optical fiber (UHNA-7) using index-matching gel to suppress parasitic reflections. The LD's emission frequency can be tuned by adjusting the driving current, and far from saturation, $\xi$ is approximately proportional to the LD injection current. 

\begin{figure}[ht]
  \centering
  \includegraphics[width=\columnwidth]{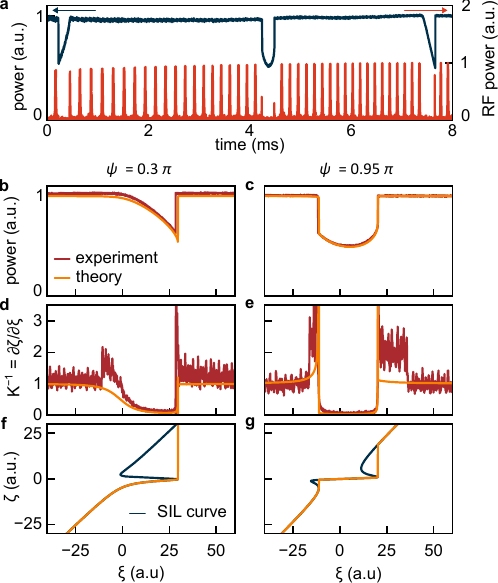}
  \caption{
      \textbf{Experimental signatures of SIL.}
      (a) Transmitted optical power (blue) during a frequency sweep of the DFB LD across three resonances of the FP microresonator and corresponding RF beatnote power (red) obtained by mixing a fraction of transmitted light with an external frequency comb with 1~GHz spacing.
      (b, c) Experimentally measured optical transmission power during the frequency sweep of the DFB LD across one FP resonator resonance for two different injection phases.
      \mbox{(d, e)} Numerically computed derivative $d\zeta / d\xi$  based on experimental data. 
      (f, g) Frequency tuning curves obtained from the theoretical model (\eqref{eq:fp_SIL}). Orange lines (retrieved from the tuning curves) in all plots indicate trajectories that the LD will follow when its injection current is linearly increased.
      }
    \label{fig:wide_sweep}
\end{figure}

In a first experiment, we sweep the DFB current with 125~mA peak-to-peak amplitude while simultaneously monitoring the transmitted power on a photodetector (see \reffig{fig:wide_sweep}{a}). As long as the LD emission frequency is far from the microresonator's eigenfrequencies, it receives next to no optical feedback from the microresonator. It is effectively free-running, with its frequency and power changing linearly. When its emission frequency approaches a resonance of the FP-resonator, a growing fraction of light is resonantly enhanced inside the FP-microresonator, reflected and injected back into the LD, resulting in SIL. This corresponds to three sharp drops in transmission, each corresponding to a resonance of the microresonator. These transmission features significantly differ from the Lorentzian resonance shape that would be observed without SIL, and all features have different shapes due to the varying injection phase, which is proportional to the frequency of the impinging light.

To calibrate the tuning behavior of the LD we also detect the beatnote between the LD's emission and an external femtosecond frequency comb of 1~GHz repetition rate. The beatnote signal is filtered by a 1.9~MHz low-pass filter, and its radio-frequency (RF) power is recorded, resulting in an RF power spike (red line, \reffig{fig:wide_sweep}{a}) each time the LD's emission frequency crosses a comb line \cite{delhaye:2009}. To improve the precision of this measurement, we also monitor the transmission of a fraction of the LD emission through an unbalanced, calibrated fiber-based Mach-Zehnder interferometer with FSR $\sim 40$~MHz. Recording such data during the LD frequency scan across a microresonator resonance enables us to derive the detunings $\xi$, $\zeta$, and therefore, the stabilization coefficient $K$ (neglecting, to a good approximation, thermal and Kerr-nonlinear resonance shift of the FP's resonance frequency).

To investigate the SIL dynamics in detail, we focus on a single FP resonance and sweep the LD frequency across it. We run this experiment for two feedback phases by adjusting the physical distance between the LD and the microresonator chip with a piezo-translation stage. The optical transmission traces are presented in \reffig{fig:wide_sweep}{(b, c)} and exhibit clear signatures of SIL. Based on the frequency comb calibration, we numerically compute $K^{- 1}=\partial{\zeta}/\partial{\xi}$ as shown in \reffig{fig:wide_sweep}{(d, e)}. Additionally, using \eqref{eq:fp_SIL} we calculate the theoretical SIL tuning curves (blue lines, \reffig{fig:wide_sweep}{(f, g)}). The orange lines in \reffig{fig:wide_sweep}{(b -- g)} present the theoretical trajectories that the LD will follow when its injection current is linearly increased, tuning the emission frequency from blue to red.
To match the theoretical predictions with the experimental data, we set the combined coupling coefficient $K_\mathrm{FP} = 44$ and optimize only the injection phase, yielding $\psi=0.3\pi$ and $0.95\pi$, respectively. We observe excellent quantitative agreement between the experimental data and the theoretical predictions. Importantly, we observe experimentally --- and confirm theoretically --- that the transmission lineshapes are unique for each phase (with a $2\pi$ period). This means they can be used to identify the injection phase. Practically, this can be useful for optimizing the SIL stabilization effect, as it is phase-dependent.

\begin{figure}[ht]
  \centering
  \includegraphics[width=\columnwidth]{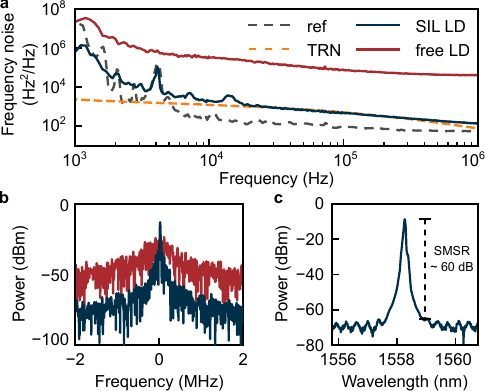}
  \caption{
      \textbf{Characterization of SIL laser}. 
      (a) Frequency noise of the free running (red) and SIL LD (blue) as well as the reference laser (gray) and the thermo-refractive noise limit simulated numerically in COMSOL (orange).
      (b) Beatnote with an external commercial laser corresponding to the respective states shown in (a).
      (c) Optical spectrum of the LD in the SIL regime.
                  }
    \label{fig:PN}
\end{figure}

Finally, we characterize the SIL laser at the feedback phase $\psi \approx \pi$, which is predicted by the theoretical model to result in the narrowest laser linewidth (i.e., the largest $K^2$, cf. Fig.~\ref{fig:tuning_curves}b). We generate a beatnote between the SIL laser and a commercial narrow-linewidth external cavity diode laser \change{(with known frequency noise)} and record the signal's quadratures with an electrical spectrum analyzer. We process this data to retrieve the SIL laser's frequency noise (FN), which is shown in \reffig{fig:PN}{a}, along with the FN of the free-running DFB LD, reference laser, and the thermorefractive noise (TRN) limit simulated numerically in COMSOL \cite{huang:2019}. We observe that the FN of the SIL laser is drastically lower than that of the free-running DFB LD and is limited by fundamental TRN noise. The corresponding radiofrequency beatnotes are shown in \reffig{fig:PN}{b}. The optical spectrum of the DFB laser in the SIL regime is shown in \reffig{fig:PN}{c}, demonstrating a side-mode suppression ratio (SMSR) greater than 60 dB.

\section{Conclusion}
In conclusion, we extend the SIL technique to chip-integrated FP microresonators, providing a novel approach for generating chip-scale, scalable, narrow linewidth lasers. We develop a theoretical model for SIL with side-coupled FP microresonators and demonstrate SIL with this resonator geometry. The side coupler enables precise and deterministic tailoring of the SIL-feedback strength. Due to high-Q and large mode volume, we observe a low-frequency noise laser, limited by fundamental TRN. As no bent waveguides are needed to form the FP microresonator, future work may use low-confinement waveguides with further increased mode volume \change{(reduction of TRN and Kerr-nonlinear effects) and higher $Q$-factor}. These results complement existing SIL techniques and are directly relevant to compact low-noise laser systems.

\begin{backmatter}
\bmsection{Funding} This project has received funding from the European Research Council (ERC) under the EU’s Horizon 2020 research and innovation program (grant agreement No 853564), from the EU’s Horizon 2020 research and innovation program (grant agreement No 101159229) and through the Helmholtz Young Investigators Group VH-NG-1404; the work was supported through the Maxwell computational resources operated at DESY.

\bmsection{Acknowledgments} We acknowledge helpful discussions with N.~Kondratiev regarding the TRN simulations and with H. Wenzel and J.-P. Koester regarding injection locking and FP cavities.

\bmsection{Disclosures}
The authors declare no conflicts of interest.

\bmsection{Data availability} Data underlying the results presented in this paper are not publicly available but may be obtained from the authors upon reasonable request.

\end{backmatter}

\bibliography{bibliography}

\end{document}